\documentclass[aps,prl,reprint,showpacs,longbibliography,floatfix,footnoteinbib]{revtex4-1}
\usepackage[latin9]{inputenc}
\usepackage{color}
\usepackage{bm}
\usepackage{amsmath}
\usepackage{amssymb}
\usepackage{graphicx}
\usepackage[unicode=true,
 bookmarks=false,
 breaklinks=false,pdfborder={0 0 1},backref=false,colorlinks=true]
 {hyperref}
\hypersetup{pdftitle={Title},
 plainpages=false,pdfpagelabels,linkcolor=blue,urlcolor=blue,citecolor=blue,pdfdisplaydoctitle=true,pdfduplex=DuplexFlipLongEdge}

\makeatletter
\usepackage{units}\usepackage{wasysym}

\definecolor{orange}{rgb}{0.50, 0.20, 0.0}

\usepackage{bm}
\usepackage{braket}

\makeatother

\begin{document}
\noindent\begin{minipage}[t]{1\columnwidth}%
\global\long\def\ket#1{\left| #1\right\rangle }%

\global\long\def\bra#1{\left\langle #1 \right|}%

\global\long\def\kket#1{\left\Vert #1\right\rangle }%

\global\long\def\bbra#1{\left\langle #1\right\Vert }%

\global\long\def\braket#1#2{\left\langle #1\right. \left| #2 \right\rangle }%

\global\long\def\bbrakket#1#2{\left\langle #1\right. \left\Vert #2\right\rangle }%

\global\long\def\av#1{\left\langle #1 \right\rangle }%

\global\long\def\tr{\text{tr}}%

\global\long\def\Tr{\text{Tr}}%

\global\long\def\pd{\partial}%

\global\long\def\im{\text{Im}}%

\global\long\def\re{\text{Re}}%

\global\long\def\sgn{\text{sgn}}%

\global\long\def\Det{\text{Det}}%

\global\long\def\abs#1{\left|#1\right|}%

\global\long\def\up{\uparrow}%

\global\long\def\down{\downarrow}%

\global\long\def\vc#1{\mathbf{#1}}%

\global\long\def\bs#1{\boldsymbol{#1}}%

\global\long\def\t#1{\text{#1}}%
\end{minipage}
\title{Fate of Quadratic Band Crossing under Quasi-Periodic Modulation}
\author{Raul Liquito$^{1}$, Miguel Gonçalves$^{1,2}$, Eduardo V. Castro$^{1,3}$}
\affiliation{$^{1}$Centro de F\'{\i}sica da Universidade do Porto, Departamento
de F\'{\i}sica e Astronomia, Faculdade de Ciências, Universidade do
Porto, 4169-007 Porto, Portugal}
\affiliation{$^{2}$CeFEMA, Instituto Superior Técnico, Universidade de Lisboa,
Av. Rovisco Pais, 1049-001 Lisboa, Portugal}
\affiliation{$^{3}$Beijing Computational Science and Research Center, Beijing
100084, China}
\begin{abstract}
We study the fate of a two-dimensional quadratic band crossing topological
phases under a one-dimensional quasi-periodic modulation. By employing
numerically exact methods, we fully characterize the phase diagram
of the model in terms of spectral, localization, and topological properties.
Unlike in the presence of regular disorder, the quadratic band crossing
is stable to the application of the quasi-periodic potential, as well
as most of the topological phase transitions occur through a gap-closing
and reopening mechanism, as in the homogeneous case. For sufficiently
high quasi-periodic potential, the quadratic band crossing point splits
into Dirac cones enabling transitions into gapped phases with Chern
numbers $C=\pm1$, absent in the homogeneous limit. This behavior
stands is in sharp contrast with the disordered case, where gapless
$C=\pm1$ phases can arise by perturbing the band crossing with any
amount of disorder. In the quasi-periodic case, we find that the $C=\pm1$
phases can only become gapless for higher potential strength. Only
in this regime do the subsequent quasi-periodic-induced topological
transitions into the trivial phase mirror the well-known ``levitation
and annihilation'' mechanism in the disordered case.
\end{abstract}
\maketitle

\section{Introduction}

\label{sec:intro}

The unique characteristics of topological insulators compared to conventional
band insulators have rendered them a central topic of current research
\citep{RevModPhys.82.3045,QZrmp11,bernevigBook,Chiu2016}. The groundbreaking
discovery of the quantum Hall effect \citep{Klitzing1980} and its
subsequent theoretical explanation through the lens of topology \citep{TKNN82,NTW85}
paved the way for the proposal of the quantum anomalous Hall effect
\citet{Haldane1988}. Remarkably, these systems can exhibit a topological
phase while no uniform magnetic field is applied and were realized
experimentally in various platforms \citep{Chang2013,Jotzu2014,Checkelsky2014,Chang2015}.
These so-called Chern insulators \citep{Chiu2016} arise from opening
non-trivial gaps in systems with band crossing points carrying a finite
quantized Berry phase, which can be accomplished by breaking time-reversal
symmetry. The simplest and more extensively studied case of a band
crossing point is that of a Dirac point, that hosts a low-energy linear
dispersion described by a Dirac Hamiltonian, as it is the case for
nodal superconductors and graphene \citep{NGPrmp}. Dirac points contain
Berry phases of $\pm\pi$, but there are other possibilities for band
crossings and associated Berry phases, as is the case of quadratic
band crossing points (QBCPs). Two-dimensional systems with QBCPs are
also of high interest because, on the one hand, they transport a finite
Berry phase of $\pm2\pi$ and on the other, a finite density of states
at QBCP, contrary to Dirac points, renders these systems unstable
to interactions \citep{SYF+09}. In QBCP systems, interactions can
induce nematic phases with two Dirac cones, each carrying half of
the QBCP's Berry phase, or gap openings that may give rise to topological
insulating phases \citep{UH11,MV14,Ray2018,Zeng2018}.

Even though topological band theory is studied in momentum-space for
translational invariant systems, topological insulators are known
to be robust to the effects of uncorrelated disorder \citet{Xiao2010}.
In fact, disorder is a key ingredient for the observation of quantized
Hall conductivity for quantum Hall systems since it localizes every
state except those responsible for the quantized Hall current that
live at very narrow energy windows \citep{KM93,OMN+03,nagaosaQSHloc07}.
In this way, varying the Fermi level in the gap filled with localized
states cannot change the Hall conductivity, giving rise to a filling-independent
plateau. Sufficiently large disorder, however, generically suppresses
the topological properties by inducing a topological phase transition
into a trivial phase, where the topological extended states meet and
become suppressed through the \textquotedblleft levitation and annihilation\textquotedblright{}
mechanism \citep{prodanBernevig,Prodan2011,nagaosaQSHloc07,Castro2015}.
However, numerous examples of disorder-induced topological phases,
known as topological Anderson insulators, have also been found \citep{Shen2009,Groth2009,Song2012,Garcia2015,Orth2016,Goncalves2018}.

A different class of systems that break translational invariance,
where more exotic localization properties can occur, are quasi-periodic
systems. Contrary to disordered systems, extended, localized and critical
multifractal phases can arise even in one-dimensional (1D) \citep{PhysRevB.43.13468,PhysRevLett.104.070601,PhysRevLett.113.236403,Liu2015,PhysRevB.91.235134,PhysRevLett.114.146601,anomScipost,GoncalvesRG2022,Goncalves_CriticalPhase}.
In higher dimensions, these systems have received considerable attention
on the interplay between moiré physics and localization \citep{Huang2016a,PhysRevLett.120.207604,Park2018,PhysRevX.7.041047,PhysRevB.100.144202,Fu2020,PhysRevB.101.235121,Wang2020,Goncalves_2022_2DMat}.
Systems with quasi-periodic modulations can be realized in widely
different platforms, including optical lattices \citep{PhysRevA.75.063404,Roati2008,Modugno_2009,Schreiber842,Luschen2018,PhysRevLett.123.070405,PhysRevLett.125.060401,PhysRevLett.126.110401,PhysRevLett.126.040603,PhysRevLett.122.170403},
photonic \citep{Lahini2009,Kraus2012,Verbin2013,PhysRevB.91.064201,Wang2020,https://doi.org/10.1002/adom.202001170,Wang2022}
and phononic \citep{PhysRevLett.122.095501,Ni2019,PhysRevLett.125.224301,PhysRevApplied.13.014023,PhysRevX.11.011016,doi:10.1063/5.0013528}
metamaterials, and moiré materials \citep{Balents2020,Andrei2020,uri2023superconductivity}.
The impact of quasi-periodic modulations on parent topological systems
has also been previously studied \citep{Fu_2021,cheng2022_december,PhysRevB.106.224505,goncalves2022topological},
and was found to give rise to interesting topological phases with
more complex localization properties than in the disordered cases.

The fate of a QBC in the presence of disorder was numerically studied
in Ref.$\,$\citep{Sobrosa2021}, in the non-interacting limit, where
it was found to be unstable to the formation of topological insulating
phases with Chern numbers $C=\pm1$, not present in the model's clean
limit. The interplay between disorder and interactions was also studied
in Refs.~\citep{Wang2017,PhysRevB.102.134204} by renormalization-group
methods, where the interaction-induced topological insulating phases
were found to be suppressed at strong disorder. The influence of quasi-periodic
modulations on QBCP has, however, remained poorly explored so far,
with the notable exception of Ref.$\,$\citep{10.21468/SciPostPhys.13.2.033}.
In this reference, a quasi-periodic potential was applied to a QBCP
system, that was found to be stable up to a flatband regime at which
the wave function becomes delocalized in momentum-space, closely related
to the incommensurate magic-angle regime found in moiré systems \citep{Fu2020,PhysRevB.101.235121,Goncalves_2021}.
The study of the topological phases that can be induced by applying
quasi-periodic modulations to a QBCP system has however remained unexplored
up to now.

\begin{figure}[!bh]
\begin{centering}
\includegraphics[width=1\columnwidth]{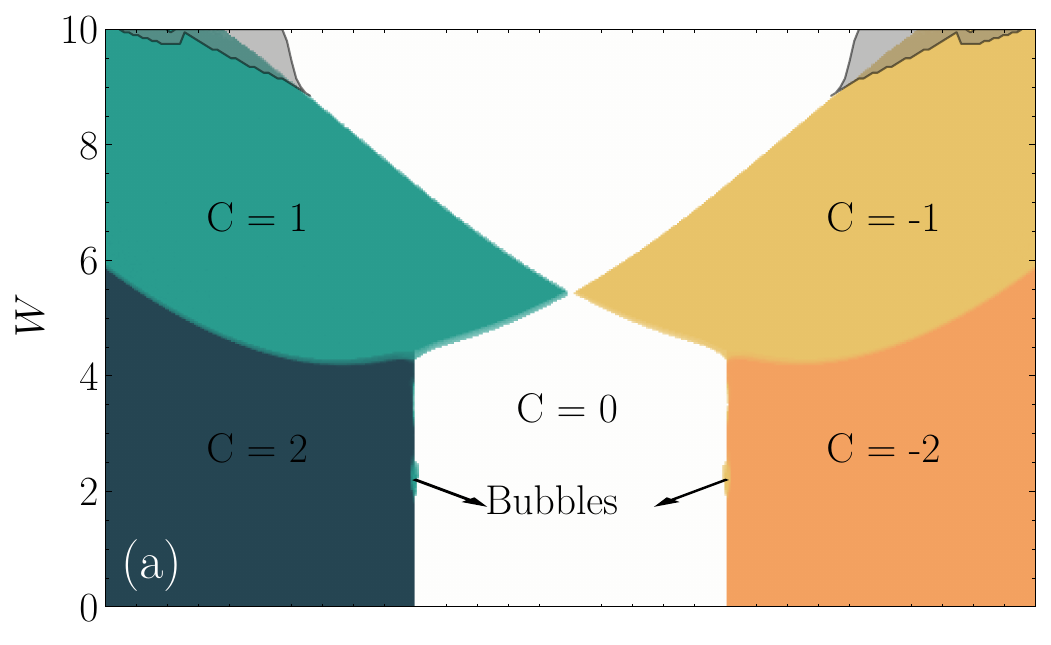}
\par\end{centering}
\begin{centering}
\includegraphics[width=1\columnwidth]{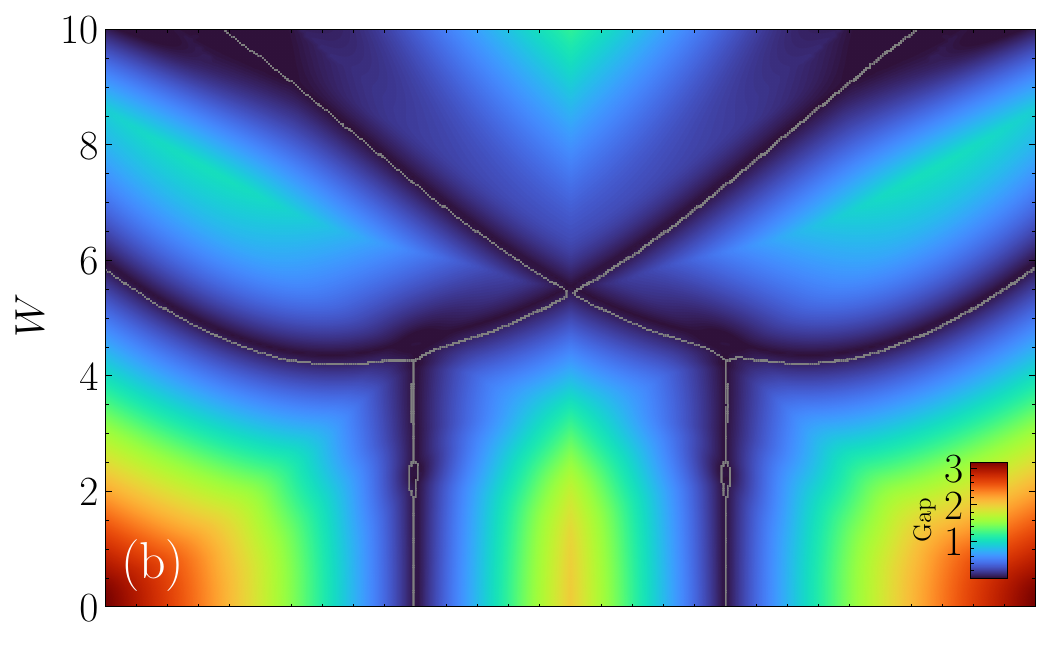}
\par\end{centering}
\begin{centering}
\includegraphics[width=1\columnwidth]{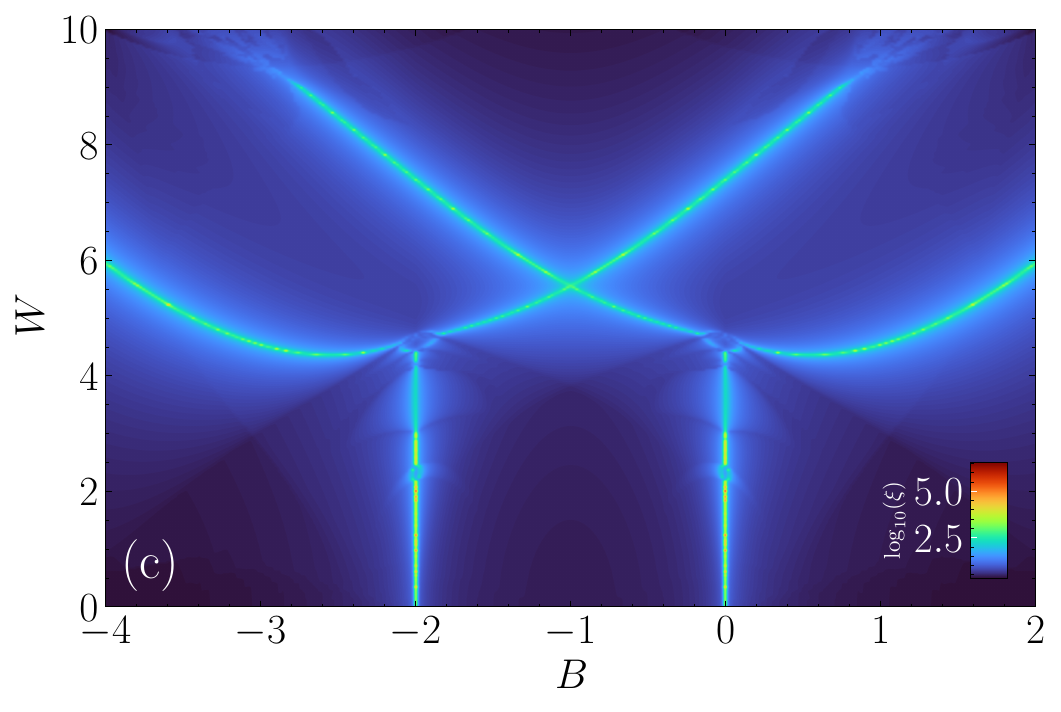}
\par\end{centering}
\centering{}\caption{\label{fig:phase-diag}(a) Chern number as a function of the model
parameter $B$ and quasi-periodic potential $W$ for a system size
of $L=21$ and $20$ different twists realizations. The greyed out
areas are gapless. (b) Energy gap as a function of $B$ and $W$ for
$L_{y}=1597$. No twist realizations were considered. The grey lines
represent the lines at which a topological phase transition occurs.
(c) Localization length ($\xi$) at the Fermi level ($E=0$) for the
$k_{x}$ values where the gap closes. A 1D system size of $N=30000$
was used. The grid discretization is $300\times300$.}
\end{figure}

In this work, we study the topological and localization properties
of a QBCP model with an applied 1D quasi-periodic modulation. The
main results are in Fig.~\ref{fig:phase-diag}. The QBCP is robust
for a small quasi-periodic potential, with the phase diagram remaining
essentially unchanged concerning the homogeneous limit, as seen in
Fig.~\ref{fig:phase-diag}(a). For higher quasi-periodic potential
strength, phases with Chern numbers $C=\pm1$, not present in the
homogeneous limit, arise. The topological transitions into these phases
occur through the gap closing and reopening mechanism, as shown in
Fig.~\ref{fig:phase-diag}(b), being therefore of a different nature
from the ones found in the disordered system in Ref.$\,$\citep{Sobrosa2021}.
Interestingly, in most of the studied topological transitions, the
gap closes at the same momentum as in the homogeneous case: at the
center and corners of the first Brillouin zone (FBZ), defined in the
translationally invariant direction. In Fig.~\ref{fig:phase-diag}(c),
the localization length at the Fermi energy, obtained by the transfer
matrix method, is seen to diverge at the gap-closing phase transitions.
Away from the Fermi energy, we found large clusters of bulk extended
states both in topological and trivial phases. This behavior contrasts
with the disordered case, where all eigenstates are localized within
the topological phases except for topologically non-trivial states
that appear at specific energies. Only for an even higher potential
do the $C=\pm1$ phases can become gapless and exhibit similar physics
to the disordered case. Further increasing the potential induces a
transition into a trivial phase that mirrors typical topological transitions
in the presence of disorder: close to the transition, topologically
non-trivial extended states arise only at very narrow energy windows,
merging at the transition, after which all states become trivial and
localized, as in the \textquotedblleft levitation and annihilation\textquotedblright{}
mechanism \citep{prodanBernevig,Prodan2011,nagaosaQSHloc07,Castro2015}.

The paper is organized as follows: In Sec.~\ref{sec:model_methods},
we introduce the tight-binding model used to describe the electronic
properties of the quasi-periodic QBC system and the methods to analyze
its properties. In Sec.~\ref{sec:results} we discuss the topological,
spectral, and localization properties. In Sec.~\ref{sec:Discussion}
we give a thorough discussion of the obtained results. We also include
three appendices: in Appendix~\ref{sec:H-matrix} we provide the
real space TB Hamiltonian and the $k_{x}$-dependent model; in Appendix~\ref{sec:Golde_ratio_search}
we present the Golden Ratio method used to compute the extrema of
the energy spectrum; finally $\text{IPR}$ and fractal dimension results
are presented in Appendix~\ref{sec:app_ipr}.

\section{Model and methods}

\label{sec:model_methods}

We consider a QBCP model on a square lattice with two orbitals per
unit cell \citep{Sobrosa2021}. This model is defined by the following
clean limit $\mathbf{k}$-space Hamiltonian

\begin{equation}
\mathcal{H}_{0}=\sum_{\mathbf{k}}\boldsymbol{\Psi}_{\bm{k}}^{\dagger}\mathcal{H_{\mathbf{k}}}\boldsymbol{\Psi}_{\bm{k}},
\end{equation}
where $\boldsymbol{\Psi}_{\bm{k}}^{\dagger}=\left(\begin{array}{cc}
c_{\mathbf{k},A}^{\dagger} & c_{\mathbf{k},B}^{\dagger}\end{array}\right)$, with $c_{\mathbf{k},\alpha}^{\dagger}$ the creation operator of
a state with crystal momentum $\mathbf{k}$ in the $\alpha$ orbital
and

\begin{equation}
\mathcal{H_{\mathbf{k}}}=\mathbf{h}(\mathbf{k})\cdot\boldsymbol{\sigma},
\end{equation}
with $\boldsymbol{\sigma}$ the vector of Pauli matrices and the vector
$\mathbf{h}(\mathbf{k})$ given by

\begin{align}
 & h_{x}=2t_{x}\sin\left(k_{x}\right)\sin\left(k_{y}\right)\nonumber \\
 & h_{y}=0\nonumber \\
 & h_{z}=2t_{z}\left(\cos\text{\ensuremath{\left(k_{x}\right)}}-\cos\left(k_{y}\right)\right).\label{eq:hvec}
\end{align}
From this point on, we set $t_{x}=t_{y}=t=1$, so that every physical
observable with energy units come in units of $t$.

QBCPs with symmetric Berry curvatures occur at $\Gamma=(0,0)$ and
$M=(\pm\pi,\pm\pi)$ points of the FBZ. For $h_{y}=0$, the system
is gapless, having a finite DOS at the Fermi level ($E=0$). The introduction
of a non-zero $h_{y}$ perturbation opens a gap that can give rise
to Chern insulating phases. This can be accomplished by introducing
the following $h_{y}$ term,

\begin{equation}
h_{y}=1+\frac{B+1}{2}\left(\cos\left(k_{x}\right)+\cos\left(k_{y}\right)\right).
\end{equation}
For this model, a single QBCP occurs at $M$ for $B=0$, and for $B=-2$
at $\Gamma$. These are the only values of $B$ for which the system
becomes gapless, where topological phase transitions between phases
with $C=2\leftrightarrow C=0$ ($B=-2$) and $C=0\leftrightarrow C=-2$
($B=0$) take place (see Fig.$\,$\ref{fig:phase-diag}(a), for $W=0$).

In this work, we will study the fate of the model's phase diagram
upon the addition of a 1D quasi-periodic potential. We consider the
Aubry-André unidirectional potential given by the real-space Hamiltonian,

\begin{equation}
\mathcal{H}_{W}=\frac{W}{2}\sum_{\mathbf{R},\alpha}\cos\left(2\pi\beta n\right)c_{\mathbf{R},\alpha}^{\dagger}c_{\mathbf{R},\alpha},\label{eq:1D_QP-potential}
\end{equation}
where $\mathbf{R}=ma\hat{\mathbf{e}}_{x}+na\hat{\mathbf{e}}_{y}$
is a lattice vector and $\beta$ is the potential frequency. In the
thermodynamic limit we choose $\beta$ to be an irrational number
in order to break translational invariance; we choose the golden ratio
$\beta=\phi_{\textrm{GR}}=(1+\sqrt{5})/2$. We carried out numerical
simulations for finite systems with $L_{x}=L_{y}=L$ (with $L$ the
number of unit cells in each direction) and periodic/twisted boundary
conditions. In order to avoid boundary defects we have chosen system
sizes $L\rightarrow L_{n}=F_{n}$, where $F_{n}$ is the $n$-th order
Fibonacci number, and approximated $\beta$ with rational approximants
of the golden ratio $\beta\rightarrow\beta_{n}=F_{n+1}/F_{n}$. This
choice ensures that the system's unit cell is of size $L$, which
guarantees that the system always remains incommensurate as $L$ is
increased. The potential in Eq.$\,$\eqref{eq:1D_QP-potential} keeps
the system translational invariant along the $x$ direction, and we
take the Fourier transform along this direction to obtain $\mathcal{H}_{k_{x}}$,
an Hamiltonian diagonal in Bloch momentum $k_{x}$ (see Appendix$\,$\ref{sec:H-matrix}).

In what follows, we carry out an extensive study of the spectral,
topological, and localization properties of the model. We compute
the Chern number phase diagram of the system via the coupling matrix
method of Ref.$\,$\citep{Fukui2005} as implemented in Ref.~\citet{Zhang2013}.
Spectral properties were studied using exact diagonalization (ED)
and the kernel polynomial method \citep{Weisse2006}.

The transfer matrix method (TMM), introduced in \citep{MK81,MacKinnon1983},
can be used to compute the localization length and, therefore, to
study bulk localization properties. At topological phase transitions,
the states are extended at the Fermi level, while in topological phases,
the system is either gapped or populated with localized states around
the Fermi level. For the two last cases, the localization length at
the Fermi level is finite, while in the former, it diverges in the
thermodynamic limit. We also used TMM to cross-check the topological
phase diagram obtained through Chern number calculations.

Throughout this work, we also realized averages over twisted boundary
conditions, such that the phase twist followed a random uniform distribution
in the interval $\theta_{i}\in[0,2\pi[$. To apply phase twists the
boundaries are periodically closed (as for periodic boundary conditions)
but with an additional phase twist, so that:

\begin{equation}
\psi_{\alpha}(\boldsymbol{R}+L\mathbf{a}_{i})=e^{i\text{\ensuremath{\theta_{i}}}}\psi_{\alpha}(\boldsymbol{R}),
\end{equation}
where $\psi_{\alpha}(\boldsymbol{R})=c_{\mathbf{R},\alpha}^{\dagger}\ket 0$.
As the system approaches thermodynamic limit, any dependence on phase
twists should vanish.

\section{Results}

\label{sec:results}

\subsection{Topological Properties}

We start by characterizing the topological phase diagram, shown in
Fig.$\,$\ref{fig:phase-diag}(a). The phase diagram was obtained
for a system with $21\times21$ unit cells averaged over $20$ random
phase twists with twisted boundary conditions.

The clean limit presents topological transitions from $C=\pm2$ to
$C=0$, with the gap closing at a QBCP ($B=\{-2,0\}$). These transitions
are robust for small $W$, still occurring at $B=\{-2,0\}$. The observed
behavior contrasts with the case of uncorrelated disorder, studied
in Ref.$\,$\citep{Sobrosa2021}, where it is shown that for infinitesimal
disorder strength $W$, $C=\pm1$ phases appear. For the present case
of an incommensurate potential, $C=\pm1$ phases only appear at large
enough potential strength, indicating a different phenomenology. Interestingly,
along the lines of fixed $B=-2,0$, when the QBCP occurs, we see the
appearance of small $C=\pm1$ bubbles, indicated by the arrows in
Fig.$\,$\ref{fig:phase-diag}(a). These bubbles are also present
in Fig.$\,$\ref{fig:phase-diag}(b) for the gap and$\,$\ref{fig:phase-diag}(c)
for the localization length, which were obtained with very different
system sizes, confirming they are not merely a result of finite-size
effects.

Increasing the quasi-periodic potential along the lines of fixed $B=\{-2,0\}$,
it will be shown below that for $W\in[3.0,4.5]$, the QBCP splits
into two Dirac cones, and, eventually, at $W\approx4.5$ a topological
phase transition into $C=\pm1$ phase occurs. For high enough potential
strength, the remaining topological phases are suppressed, and the
system transitions into a trivial phase. Unlike in the disordered
case, there are both gapped and gapless $C=\pm1$ phases separated
by the thin black lines in Fig.$\,$\ref{fig:phase-diag}(a), as we
will detail below. As we will show, these quasi-periodic-induced phases
are different from topological Anderson insulator phases that exist
for uncorrelated disorder due to their distinct localization properties.

\subsection{Gapped/Gapless Regions}

To study the spectral gap of the system, we computed the closest eigenvalue
to $E=0$, which we call $\epsilon_{0^{+}}$, using the $k_{x}$-dependent
model presented in Appendix~\ref{sec:H-matrix}. We then estimate
the gap of the system as $2\epsilon_{0^{+}}$, which is well justified
due to the particle-hole symmetry of the model, which ensures the
spectrum is symmetric around $E=0$.

In Fig.$\,$\ref{fig:phase-diag}(b), we present the gap of the system,
which was obtained using Lanczos decomposition alongside the shift
invert method for $L_{y}=1597$ at the value of $k_{x}$ where the
gap closes. Generally, we do not know the value of $k_{x}$ at which
the gap closes \footnote{At general $k_{x}$ the spectrum of $\mathcal{H}_{k_{x}}$ is gapped,
thus one might conclude that the system is gapped if the proper $k_{x}$
is not chosen}. For this reason, a golden ratio search algorithm (see Appendix.$\,$\ref{sec:Golde_ratio_search})
was implemented to compute the value of $k_{x}$ that maximizes the
valence band energy that corresponds to the gap closing point.

We can see in Fig.$\,$\ref{fig:phase-diag}(b) that the gap closes
and reopens in most topological phase transitions. We also see the
appearance of gapless regions for strong potential, seen as wider
blackish regions at high $W$. In Fig.$\,$\ref{fig:phase-diag}(a)
the greyed out regions were computed using the data in Fig.$\,$\ref{fig:phase-diag}(b).
Since the spectrum of a finite system is always discrete, we need
to properly define a threshold gap value $\delta$ below which we
consider the system to be gapless, chosen as the minimum value of
the gap at well-defined gap closing and reopening topological transitions.
For the considered system size, this value is $\delta\approx0.015$.

\subsection{Localization Properties}

To complete the characterization of the phase diagram, we now turn
to the study of the localization properties. We start by discussing
the localization length ($\xi$) at the Fermi level ($E=0$), obtained
through the TMM, and shown in Fig.~\ref{fig:phase-diag}(c).

From the gap results presented in Fig.$\,$\ref{fig:phase-diag}(b),
we know that for most of the topological phase transitions the gap
closes and reopens. Since the gap closing point should contain extended
states at a topological phase transition, the localization length
$\xi$ should diverge at these points while remaining finite in the
gapped regions. We can, therefore, capture the topological transitions
by computing $\xi$ as a function of $B$ and $W$, as shown in Fig.$\,$\ref{fig:phase-diag}(c).

Using the $k_{x}$-dependent model, we can reach closer to the thermodynamic
limit reaching system sizes of the order of $L\approx10^{4}$ with
a relative error of approximately $\epsilon\approx1\%$, where $\epsilon$
is the relative error of the average localization length after $N$
TMM iterations. To compute the localization length at the Fermi level,
we implemented TMM at the value of $k_{x}$ at which the gap closes
(calculated using the golden ratio search algorithm discussed in Appendix~\ref{sec:Golde_ratio_search}
). We have noticed that the valence band presents two local maxima,
one around the $\Gamma$-point and the other around the $M$-point,
however, the gap only closes around one of them. The TMM results in
Fig.$\,$\ref{fig:phase-diag}(c) is a sum of the localization lengths
for the two local maxima, a quantity that diverges when either of
the localization lengths does. For any other value of $k_{x}$, TMM
will always give a finite but small $\xi$ since the spectra of $\mathcal{H}_{k_{x}}$
is gapped. With this approach, we compute the localization length
$\xi$ at $E=0$ for a set of points $(B,W)$, thus obtaining in Fig.~\ref{fig:phase-diag}(c)
the contour of the original topological phase diagram shown in Fig.~\ref{fig:phase-diag}(a).
For high enough quasi-periodic potential strength, when $B\lesssim-3$
or $B\gtrsim2$, this approach fails since the gap closes and does
not reopen. The method of golden ratio search also fails in gapless
regions since the gap closes at a continuous region of $k_{x}$.

\begin{figure*}[t]
\begin{centering}
\includegraphics[scale=0.48]{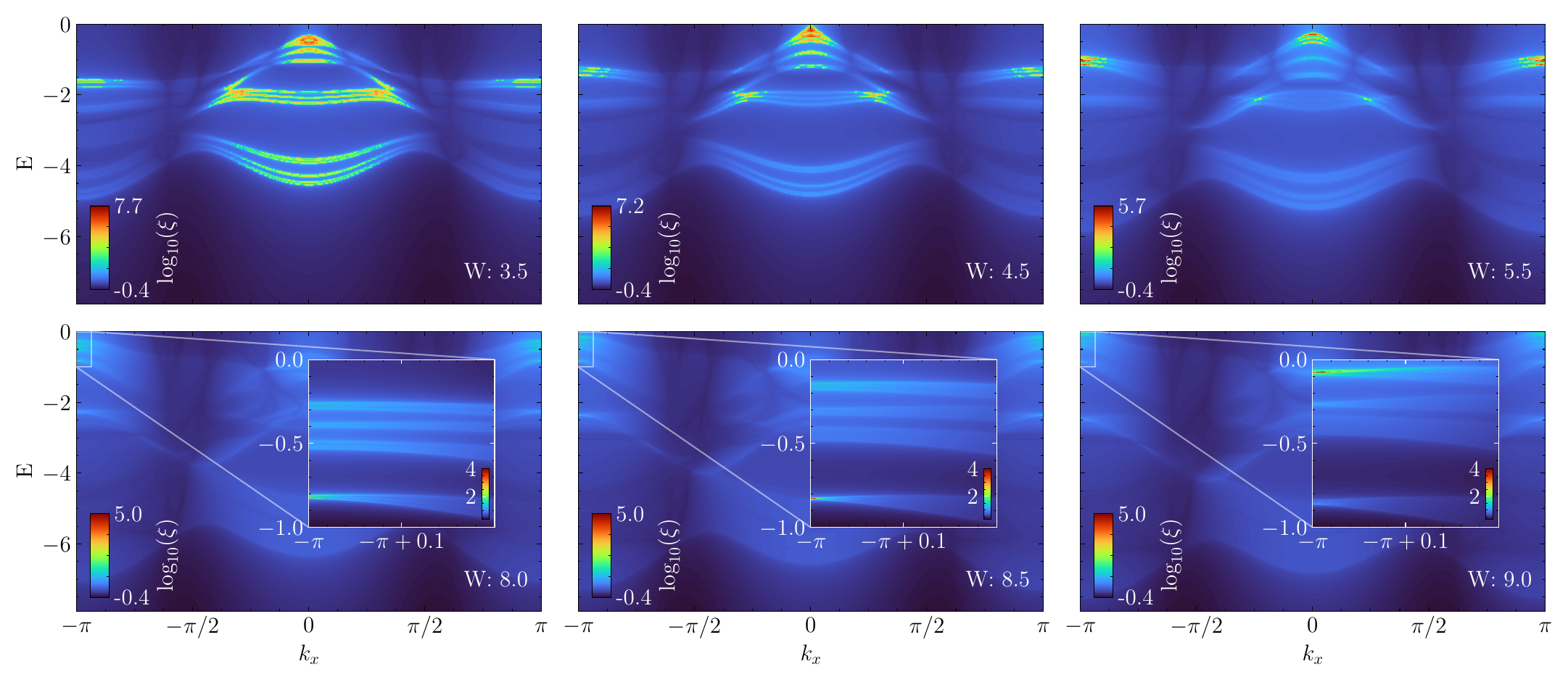}
\par\end{centering}
\centering{}\caption{Localization length as a function of energy $E$ and momentum $k_{x}$
at $B=-3$ for the entire range of the valence band. The grid discretization
is $300\times300$. The insets follow the same procedure for a smaller
region ($E\in[-1,0]$ and $k_{x}\in[-\pi,-\pi+0.2]$) with a grid
discretization of $600\times600$. \label{fig:TMM-phase-diagram-1}}
\end{figure*}

To better understand the localization properties, we have performed
an energy-resolved study considering energies away from the Fermi
level. Since localized bulk states do not contribute to the system's
topological properties, analyzing the energy position of the extended
bulk states with increasing quasi-periodic potential should give us
some intuition on how they affect the topological phases. For regular
Anderson disorder in 1D and 2D, any amount of disorder fully localizes
the spectrum for systems belonging to the orthogonal symmetry class~A
\citep{Abrahams1979}. As previously discussed, the same does not
occur in incommensurate structures for which finite fractions of extended
states can exist. To study the localization properties of the bulk
states as a function of the energy $E$ we computed the $k_{x}$ and
$E$ resolved localization length, $\xi(k_{x},E)$, for different
values of the quasi-periodic potential.

In Fig.$\,$\ref{fig:TMM-phase-diagram-1} the $k_{x}$-dependent
results for the localization length can be seen for $B=-3$. For the
first topological phase transition from $C=2$ to $C=1$ (upper panel)
there is a set of extended states around $k_{x}=0$, where the gap
closes. We can clearly see that the higher the quasi-periodic potential
$W$, the higher the fraction of localized states. On the lower panel
of Fig.$\,$\ref{fig:TMM-phase-diagram-1}, we see that right before
the second transition from $C=1$ to $C=0$, most of the bulk states
are localized except for a few extended states, which we can observe
if we use a fine enough grid discretization, as seen in the inset.
In Appendix~\ref{sec:app_ipr} we show, using inverse participation
ratio calculations, that these few extended states are really extended
and not critical. After the transition, at $W>9$ all the states are
Anderson localized, and the system is in a trivial Anderson insulating
phase.

\begin{figure*}[!t]
\begin{centering}
\includegraphics[scale=0.48]{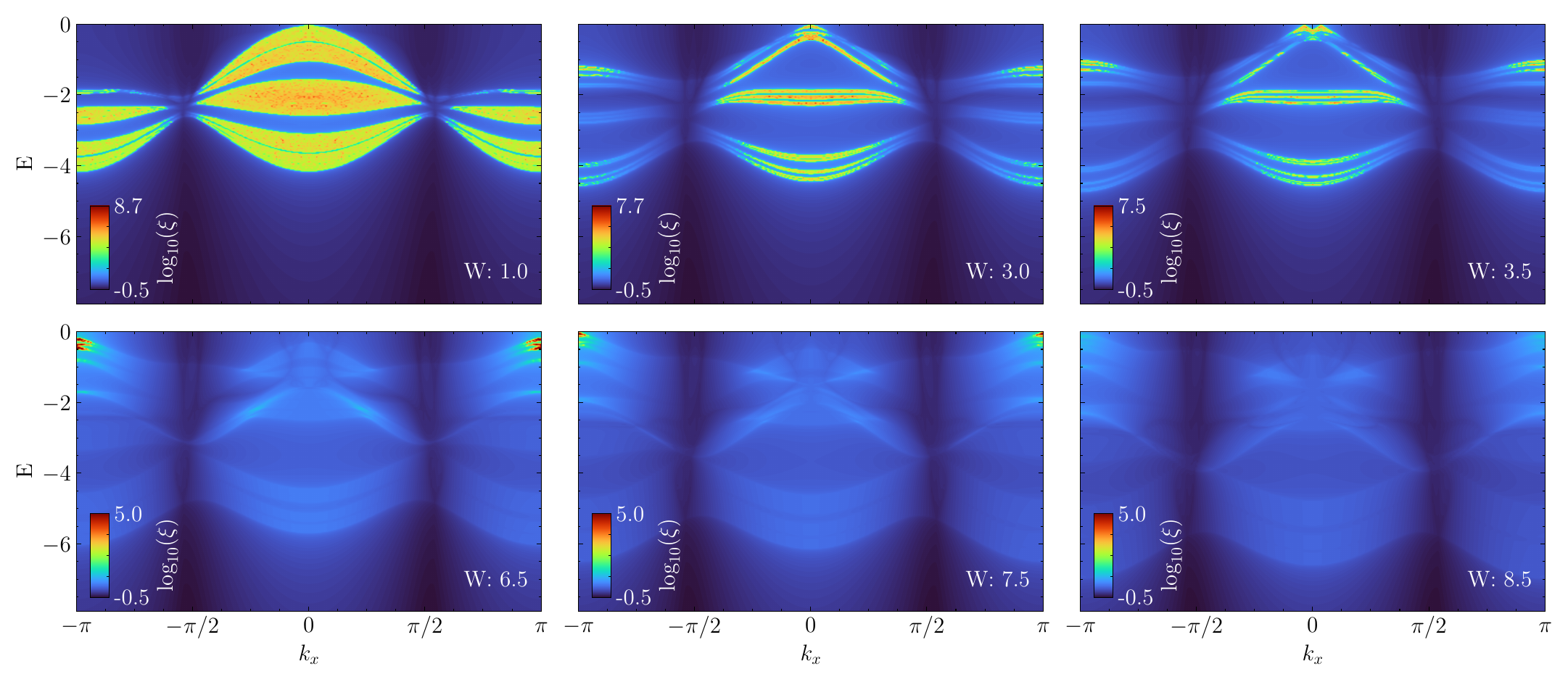}
\par\end{centering}
\centering{}\caption{Localization length as a function of energy $E$ and momentum $k_{x}$
at $B=-3$ for the entire range of the valence band. The grid discretization
is $300\times300$. \label{fig:TMM-bulk-phase-diagram_b-2}}
\end{figure*}

Figure$\,$\ref{fig:TMM-phase-diagram-1} shows the localization length
$\xi(k_{x},E)$ for $B=-2$. For small $W$ we can see that the QBCP
is robust, as expected from previous results. Around $W=3.5$, we
see that the QBCP splits into two Dirac points around the $\Gamma$
point, while under increasing quasi-periodic potential strength, bulk
states become increasingly more localized. In the lower panel of Fig.$\,$\ref{fig:TMM-phase-diagram-1},
we show the localization length results for a topological phase transition
from $C=1$ into a trivial phase. The gap closes at the transition,
after which all the states fully localize. Overall, the same behavior
follows for other values of $B$. On the other side of the diagram,
$B>-1$, the behavior is similar, with a shift of $\Delta k_{x}=\pi$.

\section{Discussion\label{sec:Discussion}}

In this work, we unveiled the complete phase diagram of a 2D system
with a QBCP under an applied 1D quasi-periodic potential. In the following,
we discuss our main results, comparing them with previous findings
for the disordered QBCP system.

In the homogeneous limit, topological phase transitions occur when
the gap closes at a QBCP. When a small quasi-periodic potential is
applied, the $y$-direction is no longer translational invariant.
Nonetheless, topological phase transitions are associated with the
gap closing in a quadratic dispersion along $k_{x}$, as seen in Figs.~\ref{fig:TMM-phase-diagram-1}
and~\ref{fig:TMM-bulk-phase-diagram_b-2}. Moreover, these transitions
occur through a gap-closing and reopening mechanism, as in the clean
case, which strongly indicates that the QBCP is stable under the addition
of quasi-periodic modulations. This behavior contrasts with the disordered
case, for which the QBCP is unstable to the appearance of gapless
phases with Chern numbers $C=\pm1$, not present in the clean limit
\citep{Sobrosa2021}. Interestingly, in the quasi-periodic case, topological
transitions into $C=\pm1$ phases were also observed for larger quasi-periodic
potentials. However, unlike the disordered case, these transitions
occur via a gap-closing and reopening mechanism. For the topological
transitions from $C=\pm2$ to $C=\pm1$, one of the QBCP is split
into two gapped Dirac cones (see, for instance, the upper right panel
in Fig.~\ref{fig:TMM-bulk-phase-diagram_b-2}), and the gap closes
only at one of them (otherwise the Chern variation would be larger
than one) \footnote{While each QBCP gives contributions $\Delta C=\pm1$ to the Chern
number, each Dirac cone only contributes with $\Delta C=\pm1/2$.
When the gap closes and reopens at a given Dirac cone, the sign of
its contribution changes, canceling with the unchanged contribution
of the remaining cone. As a result, there is a net change of $\Delta C=\pm1$
and we enter the $C=\pm1$ phases.}. In Fig.~\ref{fig:TMM-phase-diagram-1} (upper, middle panel), the
Dirac cone at a gap closing point. Once again, unlike the disordered
case, at very high potential strengths ($W\gtrsim9$), the $C=\pm1$
phases become gapless (with localized states at the Fermi level).

We also found the localization properties very distinct from the disordered
case at small quasi-periodic potential. In the presence of disorder,
most eigenstates are localized except for extended states that live
on narrow energy windows. As for the quasi-periodic case, large clusters
of bulk states remain extended. Furthermore, at high potential strengths,
an analogous of the \textquotedblleft levitation and annihilation\textquotedblright{}
mechanism can be observed close to the transitions from the $C=\pm1$
phases into the trivial phase. In fact, in this large potential limit,
the $C=\pm1$ phases can become gapless, with extended states only
existing at very narrow energy windows \footnote{Interestingly, these narrow extended states arise even slightly before
the gap closes.}. As in the disordered case, all states localize after the topological
transition into the trivial phase. Therefore, the topological and
localization properties show a mixture of features commonly observed
in the homogeneous and disordered cases, as was previously observed
for quasi-periodic Chern insulators in Ref.$\,$\citep{goncalves2022topological}.
The small potential results are also in agreement with Ref.$\,$\citep{10.21468/SciPostPhys.13.2.033},
where a different QBCP system was found to an (in this case fully
two-dimensional) quasi-periodic potential. Our findings can be verified
in ultracold atoms and trapped ions experiments, where quasi-periodic
potentials are routinely implemented \citep{PhysRevA.75.063404,Roati2008,Modugno_2009,Schreiber842,Luschen2018,PhysRevLett.123.070405,PhysRevLett.125.060401,PhysRevLett.126.110401,PhysRevLett.126.040603,PhysRevLett.122.170403}.
By keeping translational invariance along one direction, our study
of a 1D quasi-periodic potential was a suitable starting point to
address the interplay between quasi-periodicity and quadratic band
crossings. However, an equally interesting question lies on the fate
of the QBCP under the application of a fully two-dimensional quasi-periodic
potential, that we leave for future exploration.
\begin{acknowledgments}
The authors acknowledge partial support from Fundação para a Ciência
e Tecnologia (FCT-Portugal) through Grant No. UIDB/04650/2020. MG
acknowledges partial support from Fundação para a Ciência e Tecnologia
(FCT-Portugal) through Grant No. UID/CTM/04540/2019. MG acknowledges
further support from FCT-Portugal through the Grant SFRH/BD/145152/2019.
\end{acknowledgments}

\appendix

\section{Real Space and $k_{x}$-dependent Hamiltonian}

\label{sec:H-matrix}

In this section we present the $k_{x}$-dependent Hamiltonian. We
start from the Hamiltonian written in the real space basis and take
plane wave solutions along the invariant direction $x$, which enables
us to reach far closer to thermodynamic limit This transformation
comes with some caveats, since to have the full picture of the 2D
system one needs to consider the entirety FBZ along the invariant
direction, and so when computing gaps, for example, one needs to know
exactly at which $k_{x}$ does the gap close. The real space Hamiltonian
is: \ref{sec:model_methods}.

\[
\mathcal{H}=\mathcal{H}_{0}+\mathcal{H}_{B}+\mathcal{H}_{\text{1D}}
\]
Where each term is defined by:
\begin{widetext}
\begin{multline*}
\mathcal{H}_{0}=\sum_{\boldsymbol{R}}\left\{ t_{z}\left(\left|\boldsymbol{R}\pm a\hat{e}_{x},A\left\rangle \right\langle \boldsymbol{R},A\right|-\left|\boldsymbol{R}\pm a\hat{e}_{y},A\left\rangle \right\langle \boldsymbol{R},A\right|\right)\right.-\\
-t_{z}\left(\left|\boldsymbol{R}\pm a\hat{e}_{x},B\left\rangle \right\langle \boldsymbol{R},B\right|+\left|\boldsymbol{R}\pm a\hat{e}_{y},B\left\rangle \right\langle \boldsymbol{R},B\right|\right)+\\
+\frac{t_{x}}{2}\left(\left|\boldsymbol{R}\pm a\hat{e}_{x}\pm a\hat{e}_{y},A\left\rangle \right\langle \boldsymbol{R},B\right|-\left|\boldsymbol{R}\pm a\hat{e}_{x}\mp a\hat{e}_{y},A\left\rangle \right\langle \boldsymbol{R},B\right|\right)+\\
+\left.\frac{t_{x}}{2}\left(\left|\boldsymbol{R}\pm a\hat{e}_{x}\pm a\hat{e}_{y},B\left\rangle \right\langle \boldsymbol{R},A\right|-\left|\boldsymbol{R}\pm a\hat{e}_{x}\mp a\hat{e}_{y},B\left\rangle \right\langle \boldsymbol{R},A\right|\right)\right\} 
\end{multline*}

\begin{multline*}
\mathcal{H}_{B}=\sum_{\boldsymbol{R}}\left\{ i\left(\left|\boldsymbol{R},B\left\rangle \right\langle \boldsymbol{R},A\right|-\left|\boldsymbol{R},A\left\rangle \right\langle \boldsymbol{R},B\right|\right)\right.\\
-i\frac{B+1}{4}\left(\left|\boldsymbol{R}\pm a\hat{e}_{x},A\left\rangle \right\langle \boldsymbol{R},B\right|+\left|\boldsymbol{R}\pm a\hat{e}_{y},A\left\rangle \right\langle \boldsymbol{R},B\right|\right)\\
+\left.i\frac{B+1}{4}\left(\left|\boldsymbol{R},B\left\rangle \right\langle \boldsymbol{R}\pm a\hat{e}_{x},A\right|+\left|\boldsymbol{R},B\left\rangle \right\langle \boldsymbol{R}\pm a\hat{e}_{y},A\right|\right)\right\} 
\end{multline*}

\[
\mathcal{H}_{1D}=\frac{W}{2}\sum_{\mathbf{R}}\cos\left(2\pi\beta na\right)\left\{ \left|\boldsymbol{R},A\left\rangle \right\langle \boldsymbol{R},A\right|+\left|\boldsymbol{R},B\left\rangle \right\langle \boldsymbol{R},B\right|\right\} 
\]
\end{widetext}

We can now make a change of basis along the invariant direction, using
$\left|m,\alpha\right\rangle =\frac{1}{\sqrt{L_{x}}}\sum_{k_{x}}e^{-ik_{x}ma}\left|k_{x},\alpha\right\rangle $
and obtain the 1D hamiltonian $\mathcal{H}=\sum_{k_{x}}\mathcal{H}(k_{x})\left|k_{x}\left\rangle \right\langle k_{x}\right|$
\begin{widetext}
\[
\mathcal{H}(k_{x})=\mathcal{H}_{0}(k_{x})+\mathcal{H}_{B}(k_{x})+\mathcal{H}_{\text{1D}}(k_{x})
\]
where:

\begin{align*}
\mathcal{H}_{0}\left(k_{x}\right) & =\sum_{n}\left\{ t_{z}\left(2\cos\left(k_{x}a\right)\left|n,A\left\rangle \right\langle n,A\right|-\left|n\pm1,A\left\rangle \right\langle n,A\right|\right)\right.+\\
 & +t_{z}\left(-2\cos\left(k_{x}a\right)\left|n,B\left\rangle \right\langle n,B\right|+\left|n\pm1,B\left\rangle \right\langle n,B\right|\right)+\\
 & +it_{x}\sin\left(k_{x}a\right)\left(\left|n+1,A\left\rangle \right\langle n,B\right|-\left|n-1,A\left\rangle \right\langle n,B\right|\right)+\\
 & +\left.it_{x}\sin\left(k_{x}a\right)\left(\left|n+1,B\left\rangle \right\langle n,A\right|-\left|n-1,B\left\rangle \right\langle n,A\right|\right)\right\} \\
\end{align*}

\begin{align*}
\mathcal{H}_{B}(k_{x}) & =\sum_{n}\left\{ i\left(\left|n,B\left\rangle \right\langle n,A\right|-\left|n,A\left\rangle \right\langle n,B\right|\right)\right.-\\
 & -i\frac{B+1}{4}\left(2\cos\left(k_{x}a\right)\left|n,A\left\rangle \right\langle n,B\right|+\left|n\pm1,A\left\rangle \right\langle n,B\right|\right)+\\
 & +\left.i\frac{B+1}{4}\left(2\cos\left(k_{x}a\right)\left|n,B\left\rangle \right\langle n,A\right|+\left|n,B\left\rangle \right\langle n\pm1,A\right|\right)\right\} \\
\end{align*}

\[
\mathcal{H}_{\text{1D}}(k_{x})=\frac{W}{2}\sum_{n}\cos\left(2\pi a\beta n\right)\left\{ \left|n,A\left\rangle \right\langle n,A\right|+\left|n,B\left\rangle \right\langle n,B\right|\right\} 
\]
\end{widetext}

\section{Golden ratio Search Method}

\label{sec:Golde_ratio_search}

Here we present the Golden Ratio Search method, implemented to compute
the local maximum/minimum of a function of one variable $f(x)$. The
simplest implementation of the method only computes local or global
minima. However taking $f(x)\to-f(x)$ one can compute the maxima
of the function $f(x)$.
\begin{enumerate}
\item Choose two initial outside points $x_{1}$ and $x_{4}$, compute two
interior points $x_{2}=x_{4}-(x_{4}-x_{1})/z$ and $x_{3}=x_{1}+(x_{4}-x_{1})/z$
(Golden Ratio Rule), then evaluate $f(x)$ at each of the points and
set a target accuracy $\epsilon$ for the position of the minimum.
\item If $f(x_{2})<f(x_{3})$ then the minimum lies between $x_{1}$ and
$x_{3}$. Each point is shifted such that $x_{3}$ becomes the new
$x_{4}$, $x_{2}$becomes the new $x_{3}$ and the new $x_{2}$ is
computed using the Golden Ratio Rule. Evaluate $f(x)$ at the new
point.
\item Otherwise, the minimum lies between $x_{2}$ and $x_{4}$. Each point
is shifted such that $x_{2}$ becomes the new $x_{1}$, $x_{3}$becomes
the new $x_{2}$ and the new $x_{3}$ is computed using the Golden
Ratio Rule. Evaluate $f(x)$ at the new point.
\item If $x_{4}-x_{1}>\epsilon$ repeat from step $2$. Otherwise the final
estimate of the position of the minimum is $\frac{1}{2}(x_{2}+x_{3})$.
\end{enumerate}
The value of $z$ on the Golden Ratio Rule can be any value from $]0,1[$,
however the value of $z$ that lets the method converge in the least
amount of iterations is $z=\frac{1+\sqrt{5}}{2}$, the Golden Ratio.
One can now implement this algorithm to maximize $E_{-}(k_{x})$.

\section{IPR and Fractal Dimension\label{sec:app_ipr}}

Here we corroborate the results obtained via TMM at the insets of
Fig.~\ref{fig:TMM-phase-diagram-1}. We consider the IPR scaling
with system size $L$ and compute the fractal dimension $\nu$ for
the three eigenstates with bigger localization length.

\begin{figure}
\begin{centering}
\includegraphics[width=1\columnwidth]{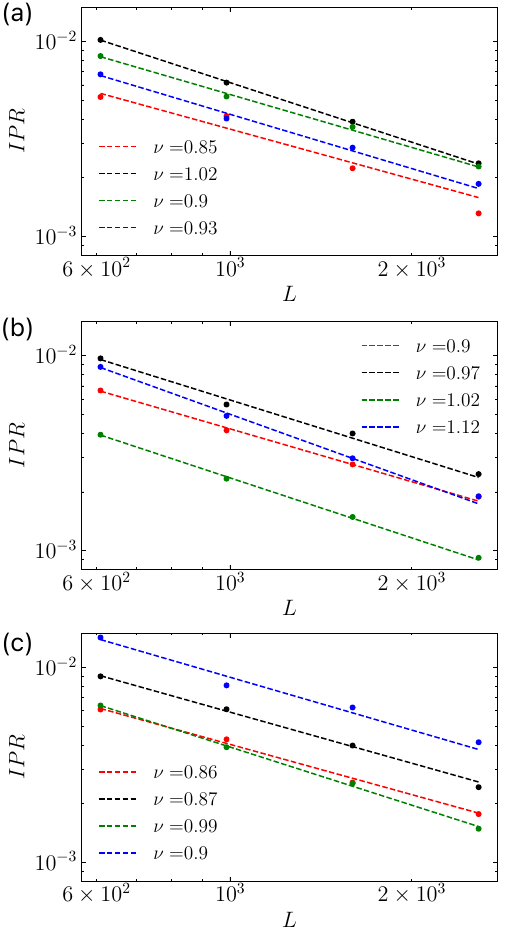}
\par\end{centering}
\caption{IPR values of the three states with the biggest localization length
obtained with TMM for $B=-3$. The fractal dimension was estimated
for each pair of $(E,k_{x})$. (a) $W=8.0$; (b) $W=8.5$; (c) $W=9.0$}
\end{figure}

We confirm that overall these states are always fully extended, yielding
$\nu\approx1$. For the states that have smaller fractal dimension
we still conclude that they are extended since $\nu$ approaches $1$
for the larger system sizes.

\bibliographystyle{apsrev4-1}
\bibliography{refs}

\end{document}